\documentclass{pasj01}

\usepackage{times}
\usepackage{url}

\begin{document}

\title{The Subaru FMOS Galaxy Redshift Survey (FastSound). I. 
Overview of the Survey Targeting on H$\alpha$ Emitters at
$z \sim 1.4$} 

\author{Motonari \textsc{Tonegawa},\altaffilmark{1}
Tomonori \textsc{Totani},\altaffilmark{1,2}
Hiroyuki \textsc{Okada},\altaffilmark{1}\\
Masayuki \textsc{Akiyama},\altaffilmark{3}
Gavin \textsc{Dalton},\altaffilmark{4,5}
Karl \textsc{Glazebrook},\altaffilmark{6}\\
Fumihide \textsc{Iwamuro},\altaffilmark{7}
Toshinori \textsc{Maihara},\altaffilmark{7}
Kouji \textsc{Ohta},\altaffilmark{7}
Ikkoh \textsc{Shimizu},\altaffilmark{1}
Naruhisa \textsc{Takato},\altaffilmark{8}
Naoyuki \textsc{Tamura},\altaffilmark{9}
Kiyoto \textsc{Yabe},\altaffilmark{10}\\
Andrew J. \textsc{Bunker},\altaffilmark{9,12}
Jean \textsc{Coupon},\altaffilmark{11}
Pedro G. \textsc{Ferreira},\altaffilmark{12}\\
Carlos S. \textsc{Frenk},\altaffilmark{13}
Tomotsugu \textsc{Goto},\altaffilmark{14}
Chiaki \textsc{Hikage},\altaffilmark{15}\\
Takashi \textsc{Ishikawa},\altaffilmark{7}
Takahiko \textsc{Matsubara},\altaffilmark{15}
Surhud \textsc{More},\altaffilmark{9}\\
Teppei \textsc{Okumura},\altaffilmark{9}
Will J. \textsc{Percival},\altaffilmark{16}
Lee R. \textsc{Spitler},\altaffilmark{17,18}\\
and Istvan \textsc{Szapudi}\altaffilmark{19}
}

\altaffiltext{1}{Department of Astronomy, School of Science, The University of Tokyo, 7-3-1 Hongo, Bunkyo-ku, Tokyo 113-0033, Japan}
\altaffiltext{2}{Research Center for the Early Universe, School of Science, The University of Tokyo, 7-3-1 Hongo, Bunkyo-ku, Tokyo 113-0033, Japan}
\altaffiltext{3}{Astronomical Institute, Faculty of Science, Tohoku University, 6-3 Aramaki, Aoba-ku, Sendai, Miyagi 980-8578, Japan}
\altaffiltext{4}{Astrophysics, Department of Physics, Keble Road, Oxford OX1 3RH, UK}
\altaffiltext{5}{RALSpace, STFC Rutherford Appleton Laboratory, HSIC, Oxford OX11 0QX, UK}
\altaffiltext{6}{Centre for Astrophysics \& Supercomputing, Swinburne University of Technology, P.O. Box 218, Hawthorn, VIC 3122, Australia}
\altaffiltext{7}{Department of Astronomy, Kyoto University, Sakyo-ku, Kyoto 606-8502, Japan}
\altaffiltext{8}{Subaru Telescope, National Astronomical Observatory of Japan, 650 North A`ohoku Pl., Hilo, Hawaii 96720, USA}
\altaffiltext{9}{Kavli Institute for the Physics and Mathematics of the 
Universe (WPI), Todai Institutes for Advanced Study,\\the University of Tokyo, 5-1-5 
Kashiwanoha, Kashiwa, Japan 277-8583}
\altaffiltext{10}{National Astronomical Observatory of Japan, Mitaka, Tokyo 181-8588, Japan} 
\altaffiltext{11}{Astronomical Observatory of the University of Geneva, ch. d' Ecogia 16, 1290 Versoix, Switzerland}
\altaffiltext{12}{Department of Physics, University of Oxford, Denys Wilkinson Building, Keble Road, Oxford, OX13RH, United Kingdom}
\altaffiltext{13}{Institute for Computational Cosmology, Department of Physics, University of Durham, South Road, Durham, DH1 3LE, UK}
\altaffiltext{14}{Institute of Astronomy, National Tsing Hua University, No. 101, Section 2, Kuang-Fu Road, Hsinchu, Taiwan 30013}
\altaffiltext{15}{Kobayashi-Maskawa Institute for the Origin of Particles and the Universe (KMI), Nagoya University, 464-8602, Japan}
\altaffiltext{16}{Institute of Cosmology and Gravitation, Portsmouth University, Dennis Sciama Building, PO1 3FX, Portsmouth, UK}
\altaffiltext{17}{Australian Astronomical Observatory, PO Box 915, North Ryde, NSW 1670, Australia}
\altaffiltext{18}{Department of Physics \& Astronomy, Macquarie University, Sydney, NSW 2109, Australia}
\altaffiltext{19}{Institute for Astronomy, University of Hawaii, 2680 Woodlawn Drive, Honolulu, HI, 96822, USA}

\email{tonegawa@astron.s.u-tokyo.ac.jp}


\maketitle

\begin{abstract}
FastSound is a galaxy redshift survey using the near-infrared Fiber
Multi-Object Spectrograph (FMOS) mounted on the Subaru Telescope,
targeting H$\alpha$ emitters at $z \sim 1.18$--$1.54$ down to the
sensitivity limit of H$\alpha$ flux $\sim 2 \times 10^{-16} \ \rm erg
\ cm^{-2} s^{-1}$.  The primary goal of the survey is to detect
redshift space distortions (RSD), to test General Relativity by
measuring the growth rate of large scale structure and to constrain
modified gravity models for the origin of the accelerated expansion of
the universe.  The target galaxies were selected based on photometric
redshifts and H$\alpha$ flux estimates calculated by fitting spectral
energy distribution (SED) models to the five optical magnitudes of the
Canada France Hawaii Telescope Legacy Survey (CFHTLS) Wide catalog.
The survey started in March 2012, and all the observations were
completed in July 2014. In total, we achieved $121$ pointings of FMOS
(each pointing has a $30$ arcmin diameter circular footprint) covering
$20.6$ deg$^2$ by tiling the four fields of the CFHTLS Wide in a
hexagonal pattern. Emission lines were detected from $\sim 4,000$ star
forming galaxies by an automatic line detection algorithm applied to
2D spectral images.  This is the first in a series of papers based on
FastSound data, and we describe the details of the survey design,
target selection, observations, data reduction, and emission line
detections.
\end{abstract}

\section{Introduction}\label{section:introduction}

High precision cosmological observations, such as the
distance-redshift relations obtained using type Ia supernovae and
baryon acoustic oscillations, the anisotropy of the cosmic microwave
background, and the large scale clustering of galaxies have
established the $\Lambda$CDM model (a universe with a non-zero
cosmological constant $\Lambda$ and cold dark matter) as the best
description of our expanding Universe (see e.g., \cite{Peebles};
\cite{Frieman}; \cite{Weinberg}; \cite{Planck}).  This result emerges
from the fact that the Universe experienced a transition from
decelerating to accelerating expansion around $z \sim 1$. The origin
of this unexpected acceleration is one of the greatest problems in
physics and astronomy. It may be a result of an exotic form of energy
with negative pressure, $\Lambda$ being one example, but the problems
of its smallness and fine tuning still remain unsolved.  Another
possibility is that the general theory of relativity (GR) is not
adequate to describe the dynamics of spacetime on cosmological
scales. If this is the case, we expect the growth rate of large scale
structure to show a deviation from the GR prediction. Therefore
measurements of the growth rate of structure at various redshifts
provides a good test of this hypothesis.

One important observable for measuring the growth rate is the redshift
space distortion (RSD) effect in galaxy redshift surveys.  Galaxy 3D
maps constructed using redshift-space distances are distorted with
respect to those in real-space due to the line-of-sight component of
the peculiar velocities of galaxies (\cite{Kaiser}).  Isotropic
statistics in real-space, such as the two-point correlation function
or the power spectrum, develop an apparent quadrupole anisotropy in
redshift space, the magnitude of which is sensitive to the velocity
power spectrum whose amplitude depends on the quantity
$f\sigma_8=d\sigma_8/d(\ln a)$.  Here $a$ is the scale factor of the
Universe and $\sigma_8$ is the rms amplitude of density fluctuations
smoothed by a top-hat filter with a comoving radius of $8h^{-1}$ Mpc.
The growth rate is well approximated by $f(z)\sim
\Omega_m(z)^{\gamma}$ where $\gamma$ is referred to as the growth
index parameter and $\Omega_m$ is the dimensionless matter density.
The value of $\gamma=0.55$ in GR, while other gravity theories predict
different values of $\gamma$ (\cite{Linder}).  Comparison between the
observed value of $f \sigma_8$ and its theoretical prediction at
various redshifts is a good test of GR on cosmological scales
(\cite{Song}), assuming that galaxy biasing is linear.  This test has
already been performed at $z<1$ using data from a number of galaxy
redshift surveys (\cite{Hawkins}; \cite{Guzzo}; \cite{Blake};
\cite{Samushia}; \cite{Reid}; \cite{Beutler2012}; \cite{delaTorre};
\cite{Beutler2014}). The standard $\Lambda$CDM model is found to be
consistent with the observed values of $f \sigma_8$ and its evolution.

A next important step for this test is to reduce the statistical error
on $f \sigma_8$ using larger galaxy surveys at $z < 1$, or to extend
the $f \sigma_8$ measurements to even higher redshifts.  At higher
redshifts, nonlinearities on the physical scales of interest are
smaller than they are today, which can result in a cleaner measurement
of $f \sigma_8$. Moreover, at high redshifts, $f\sim1$, and RSD is
directly sensitive to $\sigma_8$, thus providing a baseline for lower
redshifts measurements.  RSD has been detected at $z\sim3$ using
Lyman-break galaxies (\cite{Bielby}), at the significance of
$\sim2\sigma$ level.  The principal aim of the FastSound\footnote{The
  name is composed of two acronyms: one a Japanese name ``FMOS Ankoku
  Sekai Tansa'' (FMOS ˆÃ•¢ŠE'T¸, meaning FMOS dark Universe
  survey), and the other an English name ``Subaru Observation for
  Understanding the Nature of Dark energy''.} galaxy redshift survey
is to measure $f \sigma_8$ at $z \sim$ $1.18$--$1.54$ for the first
time by detecting RSD using the near-infrared Fiber Multi-Object
Spectrograph (FMOS) mounted on Subaru Telescope.

The FMOS instrument has $400$ fibers in a $30$-arcmin diameter
field-of-view, and covers the wavelength range $0.9$--$1.8 \ \mu$m in
its low-dispersion mode; for FastSound we use the higher-throughput
higher dispersion mode working in the $H$-band and covering
$1.43$--$1.67\mu$m.  This allows us to reach the $z>1$ universe by
detecting bright H$\alpha$ $\lambda6563$ emission lines from
star-forming galaxies.  FMOS uses a novel fiber positioning system,
called ``Echidna'', which is driven by a saw-tooth voltage pulse sent
to the piezoelectric actuator, in order to populate $400$ fibers in a
limited space of $15$ cm diameter.  The atmospheric OH emission lines
are hardware-blocked using an OH mask mirror, thus eliminating a
dominant noise source in the near-infrared region.  

All the FastSound observations have been completed as the Subaru
Strategic Program for FMOS (PI: T. Totani), using $35$ nights from
March 2012 to July 2014. About $4,000$ emission line galaxies were
detected with $S/N\geq4.0$ in the total survey area of $20.6$ deg$^2$
[$121$ FMOS field-of-views (FoVs)], and more than $90\%$ of them are
considered to be H$\alpha$ emitters with a median redshift of
$z\sim1.4$.  It is expected that RSD will be detected at $\sim4\sigma$
level from this data set (Paper IV, Okumura et al., in preparation).
In addition to enabling a RSD measurement for cosmology, this sample
will also provide fresh insights into the evolution of the star
formation and metallicity of galaxies at intermediate redshift.

In this paper, we present an overview of FastSound, especially
focusing on the survey design, the details of observations and data
reductions. We describe the observed survey fields in
\S\ref{section:fields}, and the target selection based on photometric
redshift and H$\alpha$ flux estimates in \S\ref{subsection:selection}.
In \S\ref{section:observation}, we present details of our FMOS
spectroscopic observations.  This is followed in
\S\ref{section:reduction} by a description of the data processing
using the standard FMOS data reduction pipeline (FIBER-pac;
\cite{Iwamuro}) and automated emission line search by the software
FIELD (\cite{Tonegawa2}).  Finally, we present a brief summary in
\S\ref{section:conclusion}.  Throughout this paper, all magnitudes are
given in the AB system, and coordinates in the equinox J2000.0
system. We adopt a standard set of the cosmological parameters:
$(\Omega_m,\Omega_\Lambda,h) = (0.3,0.7,0.7)$, where $h$ is
$H_0/(100\,{\rm km/s/Mpc})$.

\section{Survey Fields}\label{section:fields}
\begin{figure*}
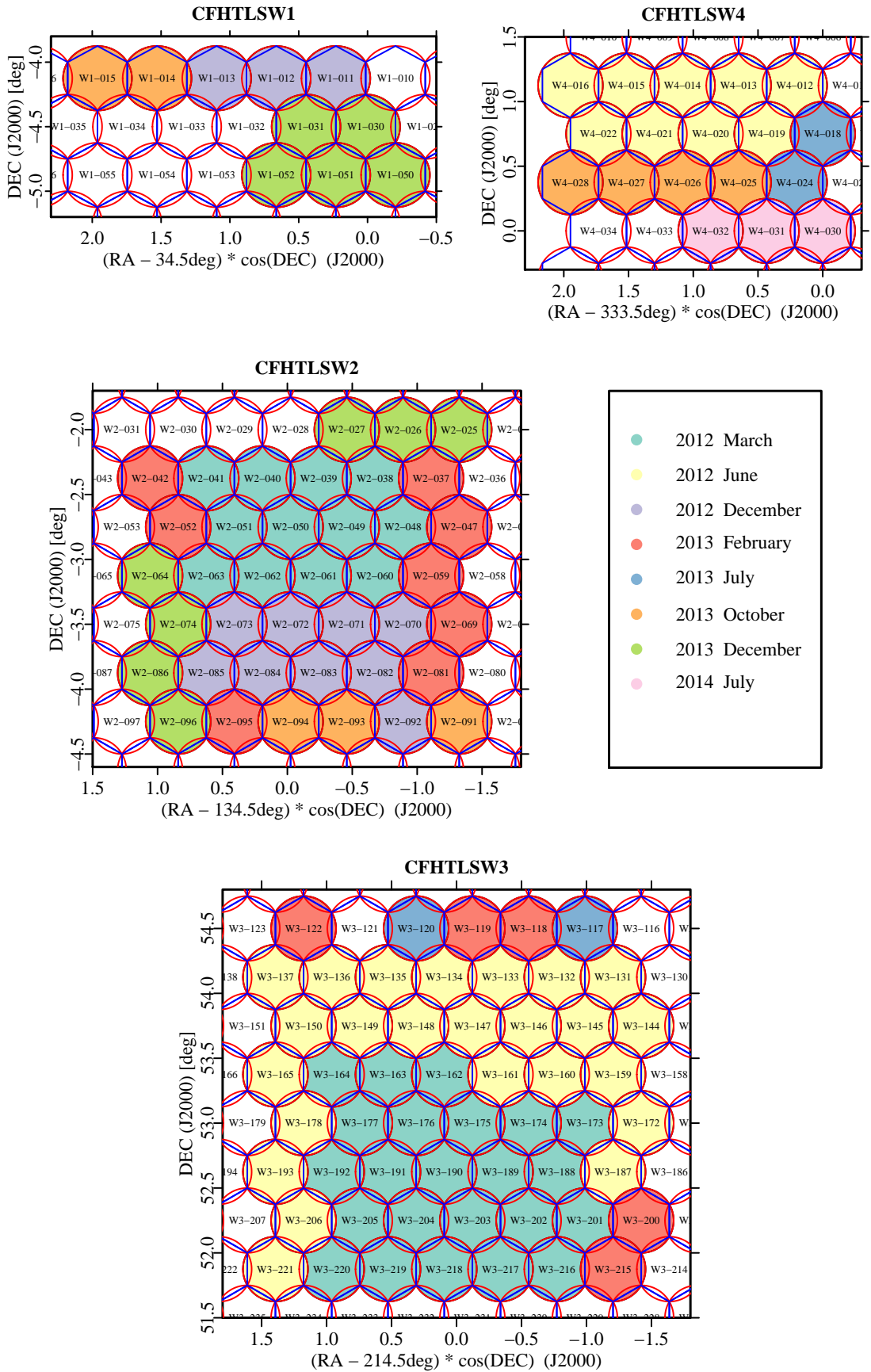

 \begin{center}
 \FigureFile(78.7mm,48.3mm){figure1a.eps}
 \FigureFile(70.6mm,56.7mm){figure1b.eps}
 \FigureFile(85.8mm,81.0mm){figure1c.eps}
 \FigureFile(49.5mm,81.0mm){figure1d.eps}
 \FigureFile(92.5mm,89.7mm){figure1e.eps}
 \end{center}
 \caption{The footprints of the observed FastSound field-of-views
   (FoVs).  FMOS FoVs (shown by red circles, 30 arcmin diameter) are
   arranged in a continuous hexagonal tiling.  The FMOS FoV field IDs
   are indicated in each FoV.  FoVs actually observed by FastSound are
   filled by different colors corresponding to different observing
   periods, as indicated in the figure, while the white FoVs were not
   observed.  }\label{figure:fov}
\end{figure*}

The spectroscopic targets for FastSound were selected from the four
fields (W1--W4) of the Canada-France-Hawaii Telescope Legacy Survey
(CFHTLS) Wide Fields which cover an area of $\sim 170$ deg$^2$ in
total with five optical band filters of $u^*, g', r', i', z'$
(\cite{Goranova}). FastSound spans a total of $20.6$ deg$^2$ in these
fields.  Figure \ref{figure:fov} shows the FastSound footprints in the
four fields.  The FastSound observation regions were chosen so as to
maximize its legacy value (i.e., overlap with existing data).  In
CFHTLS W1, the survey data of HiZELS (\cite{Geach}; \cite{Sobral}),
SXDS (\cite{Furusawa}), UKIDSS-DXS and UDS (\cite{Lawrence}), VVDS
Deep (\cite{Febre}) and Wide (\cite{Garilli}), VIPERS
(\cite{Guzzo14}), and CFHTLS Deep 1 Field overlap with FastSound.
Similarly, the FastSound region in W3 overlaps with the DEEP3
(\cite{Cooper}) field, and that in W4 overlaps with the UKIDSS-DXS and
VVDS Wide fields. Such overlaps would be useful to better understand
the physical properties of emission line galaxies detected by
FastSound [e.g., stellar mass from NIR data, see Paper III (\cite{Yabe})
for a study of the mass-metallicity relation].

For FastSound, we adopted a gapless tiling pattern of hexagons
inscribed in the FMOS circular FoV of $30$ arcmin diameter, as
illustrated in Figure \ref{figure:fov}.  The differences of RA and DEC
between two neighboring hexagons satisfy $\Delta$(RA) $\cos$(DEC) =
$\sqrt{3}r$ and $\Delta$(DEC) = $(3/2)r$, where $r = 0.25$ deg is the
radius of the FMOS FoV.  The FastSound FoV-ID is related to the
central point coordinates by the running integers $i$ and $j$ such
that
\begin{equation}
{\rm FoV \ ID \ number} = A_k i + j + B_k
\end{equation}
\begin{equation}
{\rm DEC} = d_k - \frac{3}{2} r i
\end{equation}
\begin{equation}
{\rm RA'} = r_{1,k} + \sqrt{3} r j \pm \frac{\sqrt{3}r}{2}\delta_i \ \ \mbox{\small ($+$:W1,W4, $-$:W2,W3)}
\end{equation}
\begin{equation}
{\rm RA} = r_{2,k} + \frac{ {\rm RA'} }{ \cos ({\rm DEC}) }
\end{equation}
where $\delta_i = $ 0 (for even $i$) or 1 (for odd $i$), $A_k$ and
$B_k$ are integers, while $d_k$, $r_{1,k}$, and $r_{2,k}$ are
non-integers.  These are defined in Table \ref{table:fov_param} for
the $k$-th field of CFHTLS-W ($k = $ 1--4). The ranges of $i$ and $j$
are also given in the Table.  Because $A_k$ means the number of grids in $j$
(i.e., $j_{\max, k} - j_{\min, k} + 1$), one can calculate
RA and DEC if the field ID number is given, by finding $i$ and $j$ as
the quotient and remainder of (ID $ - B_k$) divided by $A_k$.  The
FastSound FMOS FoVs are then specified by the field ID (W1--4) and the
FoV ID; for example, the W1\_030 FoV has $i=1$ and $j=9$, and its
central coordinates are (RA, DEC) = (34.5137, $-$4.5000) in degrees.

\begin{table*}
\begin{center}
\caption{The parameters relating the FMOS FoV ID numbers and 
the central coordinates}\label{table:fov_param}
\begin{tabular}{cccccccccc}
\hline
\hline
$k$ & $A_k$ & $B_k$ & $d_k$ & $r_{1,k}$ & $r_{2,k}$ & $i_{\min}$ & $i_{\max}$ & $j_{\min}$ & $j_{\max}$ \\
\hline
$1$ & $20$ & $1$ & $-4.125$ & $-4.1$ & $34.5$ & $0$ & $14$ & $0$ & $19$\\
$2$ & $11$ & $1$ & $-1.25$ & $-2.19$ & $134.5$ & $0$ & $11$ & $0$ & $10$\\
$3$ & $14$ & $1$ & $57.5$ & $-2.72$ & $214.5$ & $0$ & $16$ & $0$ & $13$\\
$4$ & $6$ & $6$ & $1.5$ & $0.0$ & $333.5$ & $0$ & $6$ & $-1$ & $4$\\
\hline
\hline 
\multicolumn{10}{l}{
Note: $d_k$, $r_{1,k}$, and $r_{2,k}$ are in units of degrees.}
\end{tabular}
\end{center}
\end{table*}

The survey was originally planned to cover approximately the same area
for the four fields of W1--4, but because of weather conditions and/or
telescope/instrument problems, there is a considerable difference in
the areas actually observed in each of the fields: $10$, $39$, $54$, and $18$ FoVs in the W1--4 fields,
respectively. The footprints of FastSound observations are indicated
in Figure \ref{figure:fov}, with different colors for different
FastSound observing runs.

\begin{figure}
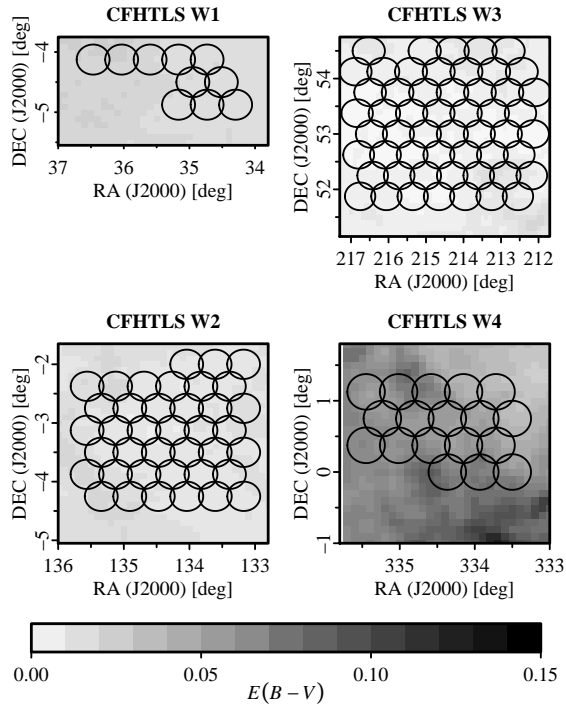

 \begin{center}
 \FigureFile(75mm,94mm){figure2.eps}
 \end{center}
 \caption{The Galactic extinction map [$E(B-V)$ from \cite{Schlegel}]
   in each of the four FastSound survey fields. }
 \label{figure:ebv}
\end{figure}

\section{Target Selection}\label{subsection:selection}

After an examination of various selection methods for FastSound
(\cite{Tonegawa}), we chose the photometric redshift and star
formation rate estimate based on SED fittings to the five CFHTLS-W
optical bands as the FastSound target selection criteria.  We started
from the $z$-band selected CFHTLS galaxy catalog produced by Gwyn
(2012).  We used the MAG\_AUTO magnitudes, and the limiting magnitudes
are $26.0$, $26.5$, $25.9$, $25.7$, and $24.6$ for $u^*g'r'i'z'$
respectively at $50\%$ completeness level for point sources
(\cite{Gwyn}). Magnitudes corrected for the Galactic extinction using
the $E(B-V)$ map of \citet{Schlegel} were used in the following
procedures of target selection.  The extinction maps in the FastSound
survey regions are displayed in Figure \ref{figure:ebv}. The W4 field
has higher extinction ($E(B-V)=0.065$ on average) than the other
fields ($E(B-V)=0.012$--$0.024$ on average).

We used the official CFHTLS photometric redshifts ($z_{\rm ph}$) provided
for galaxies brighter than $i'=24.0$ (\cite{Ilbert}; \cite{Coupon})
\footnote{\url{http://terapix.iap.fr/}}. We also considered galaxies
for which official $z_{\rm ph}$'s are not provided, as potential
FastSound targets.  We determined the photometric redshifts of these
galaxies in the following manner.  We used the public code
\textit{LePhare} (\cite{Arnouts}; \cite{Ilbert}), with the same
templates and parameters as those for the CFHTLS official photo-$z$
calculation.  Photo-$z$ training (i.e., zero-point correction for the
input photometry) was performed using VVDS Deep and Wide data
(\cite{Febre}; \cite{Garilli}). {\it LePhare} provides the best-fit
(minimum $\chi^2$) $z_{\rm min}$ and the median of likelihood
distribution $z_{\rm med}$ as estimates of $z_{\rm ph}$.  While
$z_{\rm med}$ was used as the CFHTLS official photometric redshift,
$z_{\rm min}$ was adopted for galaxies calculated by us.  (We noticed
this difference after the survey started.)  Galaxies whose photo-$z$'s
were calculated by us are relatively faint compared with those with
the official CFHTLS photo-$z$'s, comprising 4.7\% of all the final
FastSound target galaxies and 1.9\% of the emission line galaxies
detected.  The latter fraction is lower than the former, probably
because emission lines of fainter galaxies in $i'$ are weaker and
hence more difficult to be detected by FMOS.

There is a tight correlation between star formation rate (SFR) and
H$\alpha$, and H$\alpha$ is widely recognized as the best star
formation rate indicator of galaxies (\cite{Kennicutt}). Therefore
estimates of SFR are also important for efficiently selecting
H$\alpha$ emission line galaxies.  However, the SED templates used in
the photo-$z$ calculation are empirical ones, and hence cannot be used
to infer a SFR. Therefore we have also performed SED fitting with the
theoretical PEGASE2 templates (\cite{Fioc}) assuming the Scalo IMF
(\cite{Scalo}) and the solar metallicity, again using
\textit{LePhare}, with redshifts fixed at $z_{\rm ph}$ calculated
using the empirical templates. Model templates have exponential star
formation histories with decaying timescales of $0.1$--$20$ Gyr and
ages of galaxies in the range of $0.3$--$10$ Gyr.  The extinction is
assumed to be the Calzetti law (\cite{Calzetti}). We then estimated
H$\alpha$ luminosities from SFR and $E(B-V)$ calculated by {\it
  LePhare} (best-fit values at the $\chi^2$ minimum), as
\begin{eqnarray}
\log{(L_{\rm H\alpha}/[{\rm erg/s}])} &=& 
40.93 + \log{({\rm SFR/[M_\odot/yr]})} \nonumber \\
 && - 0.4 A_{\rm H\alpha}.
\end{eqnarray}
The conversion factor $40.93$ was derived by comparing the spectroscopic
H$\alpha$ line luminosities of the SDSS galaxies with those estimated
by photometric fittings, using the method of Sumiyoshi et al. (2009).
The value is different from that of Kennicutt (1998) and Sumiyoshi et
al. (2009), because of the use of the Scalo IMF rather than the
Salpeter IMF.  Here, extinction of H$\alpha$ photons is expected to be
larger than that for stellar radiation, and $A_{\rm H\alpha}$ is
related to stellar $A_V$ using the prescription discussed in Cid
Fernandes et al. (2005).  The stellar $A_V$ is estimated from $E(B-V)$
obtained from the SED fitting, assuming the Calzetti law again.
Finally $\log{L_{\rm H\alpha}}$ is converted to H$\alpha$ flux
assuming the photometric redshifts.

We selected galaxies at $1.1 < z_{\rm ph} < 1.6$, according to the
redshift range of H$\alpha$ ($1.18 \le z \le 1.54$) corresponding to
the observing wavelength range of FastSound ($1.43$--$1.67 \ \mu$m
using the high resolution mode of FMOS).  The range $1.1<z_{\rm
  ph}<1.6$ is wider than the wavelength coverage of FMOS, to allow the
$z_{\rm ph}$ range to span the wavelength coverage, including the
uncertainties in $z_{\rm ph}$. The typical error, $\sigma_{\Delta
  z/(1+z_s)}\equiv 1.48\times \mathrm{median}(\left|\Delta
z\right|/(1+z_s))$, is $0.08$ for galaxies in our pilot observing run
(\cite{Tonegawa}).

\begin{figure*}
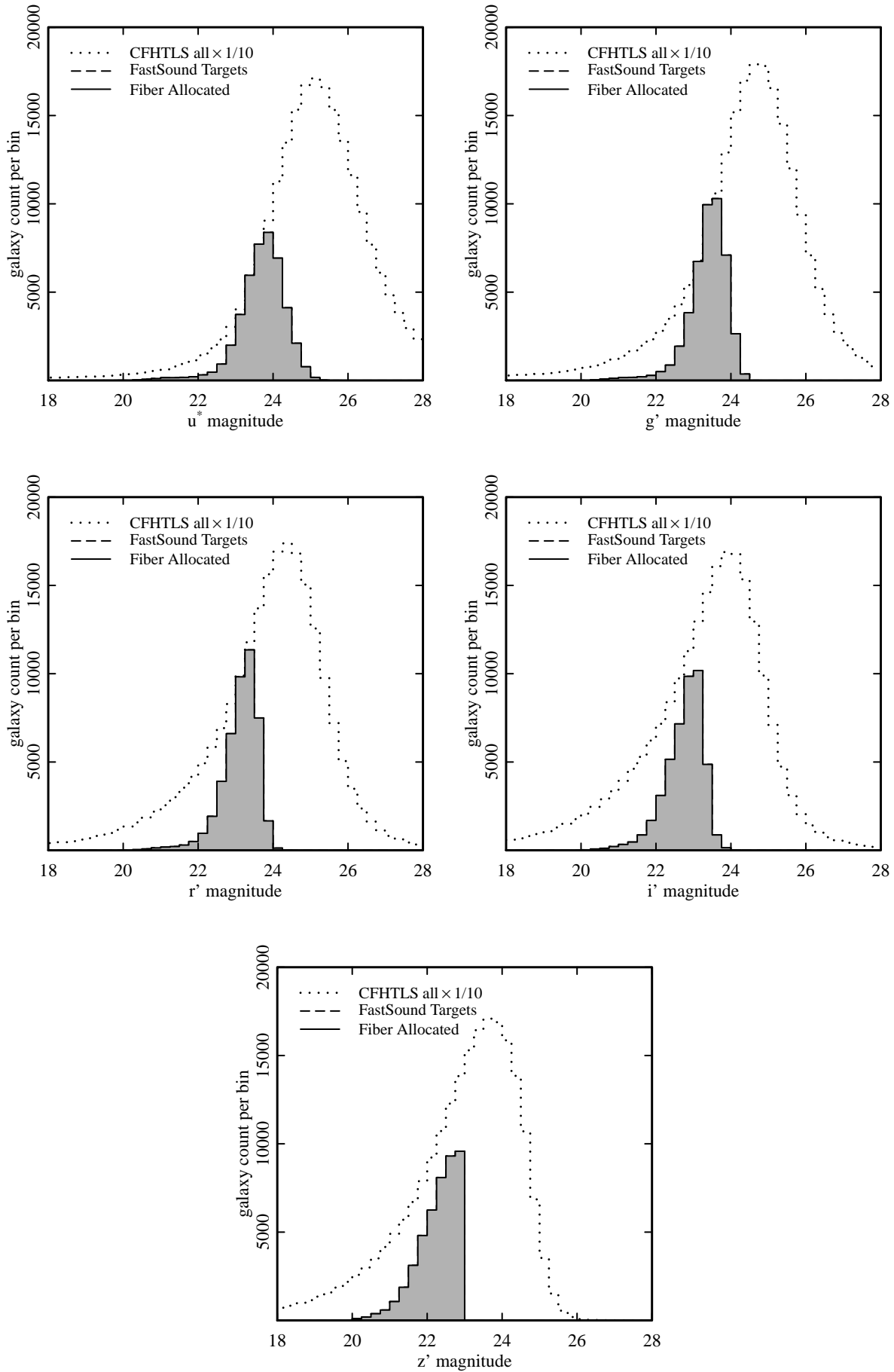

 \begin{center}
 \FigureFile(80mm,80mm){figure3a.eps}
 \FigureFile(80mm,80mm){figure3b.eps}
 \FigureFile(80mm,80mm){figure3c.eps}
 \FigureFile(80mm,80mm){figure3d.eps}
 \FigureFile(80mm,80mm){figure3e.eps}
 \end{center}
 \caption{The histogram of $u^*g'r'i'z$ magnitudes of all CFHTLS Wide
   galaxies in the FastSound footprints (dotted), FastSound target
   galaxies (dashed), and galaxies that the fibers were actually
   allocated to (solid). The cutoff of $z'$-magnitude at $23.0$ is due
   to our selection condition.
}
\label{figure:magnitudes}
\end{figure*}

We further selected galaxies satisfying $20.0 < z' < 23.0$ and $g' -
r' < 0.55$; these conditions were introduced to increase the success
rate of emission line detection by FMOS, based on past experience (see
\cite{Tonegawa}).  Finally, we set a threshold for the estimated
H$\alpha$ flux, to select a required number of H$\alpha$ bright
galaxies.  In FastSound, each FMOS FoV is visited only once with one
fiber configuration, and hence the threshold flux is set so that the
number of target galaxies within FMOS FoV is about $\sim500$, slightly
more than the number of FMOS fibers, allowing for dropouts due to the
fiber collision or the density contrast of galaxies. A fixed threshold
flux is adopted within one of the four CFHTLS Wide fields, but it is
different for different fields: $1.0\times10^{-16}\;{\rm erg \ cm^{-1}
  \ s^{-1}}$ for W1/W2, $1.1\times10^{-16}\;{\rm erg \ cm^{-1}
  \ s^{-1}}$ for W3, and $0.9\times10^{-16}\;{\rm erg \ cm^{-1}
  \ s^{-1}}$ for W4.  This final galaxy sample within each FMOS FoV is
sent to the fiber allocation software of FMOS.  (In this work we call
this sample as the target galaxies in FastSound, which is different
from the galaxy sample that were actually observed by FMOS, because
fiber allocation is not complete. See \S\ref{subsection:fiber_allocation}.) 
The $u^*g'r'i'z'$ magnitude distribution of the target galaxies and
fiber-allocated galaxies compared to all CFHTLS-Wide objects in the survey
footprint is displayed in Figure \ref{figure:magnitudes}.  Our target
selection is biased towards blue star-forming galaxies, but this is
not a serious problem for $f\sigma_8$ measurement within the linear
regime, as mentioned in \S\ref{section:introduction}.

\section{Observations}\label{section:observation}

\subsection{Basic Strategy}
\label{section:basic_strategy}

FMOS has two modes for fiber configuration: the normal-beam switch
(NBS) mode and the cross-beam switch (CBS) mode. In NBS mode, all the
$400$ fibers are allocated to target objects in an on-source exposure,
and then the FoV is offset to observe the off-source sky background,
with the same fiber configuration.  In CBS, on the other hand, the
fibers are split into two groups of $200$ fibers, and one observes
targets while the other observes the sky background. Two exposures are
taken for one target set, exchanging the role of two fiber groups.
The CBS mode has the advantage of shorter observing time required to
achieve a fixed $S/N$ when targets are less than $200$ in a
FoV. However, CBS has the disadvantage of more complicated fiber
configuration, resulting in a smaller number of allocated spines than
NBS. Therefore we chose NBS for the FastSound project.

FMOS provides two different spectral resolution modes: the
low-resolution (LR) mode and the high-resolution (HR) mode.  The LR
mode covers $0.9$--$1.8 \ \mu$m with a typical spectral resolution of
$R\sim500$, while HR covers a quarter of the wavelength range covered
by LR with $R\sim2200$ where $R=\lambda/\Delta \lambda$ and $\Delta
\lambda$ denotes FWHM.  The HR mode was chosen for FastSound because
in LR the additional volume-phase holographic (VPH) grating used to
anti-disperse spectra decreases the instrumental throughput by a
factor of about two (\cite{Kimura}).  The wavelength range was set to
be $\lambda=1.43$--$1.67$ $\mu$m (called H short+), corresponding to
H$\alpha$ $\lambda6563$ at $z = 1.18$--$1.54$.  In the H short+
wavelength range, the spectral resolution changes within the range of
$R \sim 2000$--$2700$, with the typical value of $2400$. The detector
size is 2k$\times$2k, and hence the pixel scale is $\sim1.1\;{\rm
  \AA/pix}$.

The FastSound survey observed each FMOS FoV with a single set of fiber
configuration including $\sim400$ targets, selected by the procedures
described in the previous section.  The light from the targets
collected by the $400$ fibers of FMOS is sent to the two
spectrographs, IRS1 and IRS2, each of which produces about $200$
spectra.  For each FoV we observed two sets of $15$ min on-source and
$15$ min off-source, i.e., total $30$ min on source.  Including $30$
min of overhead, each FoV typically took about $90$ min, allowing us
to observe about $6$ FoVs per night. Sometimes three or more source
frames were taken in one FoV, depending on the quality of the data
affected by weather conditions.  

\begin{figure}
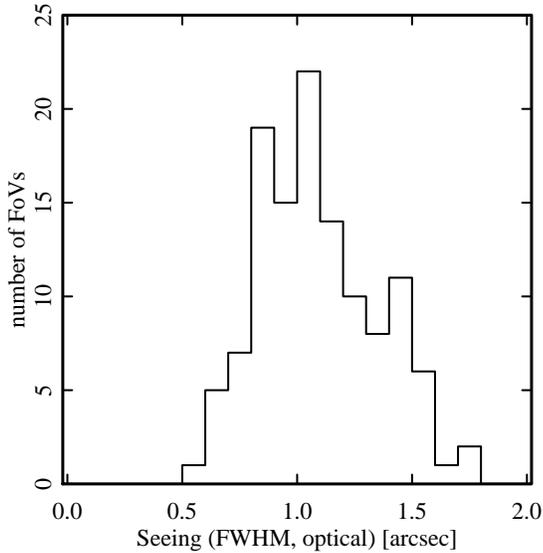

 \begin{center}
 \FigureFile(80mm,80mm){figure4.eps}
 \end{center}
 \caption{The histogram of seeing (FWHM) for all the FMOS FoVs in the
   FastSound survey.}
\label{figure:seeing}
\end{figure}

Seeing was measured during the observing time for each FoV, using an
optical image of coordinate calibration stars (see
\S\ref{subsection:fiber_allocation}) taken by the sky camera (CCD
camera in the Echidna Focal Plane Imager).  The seeing in the
FastSound data ranges from $\sim 0.6$ to $1.6\;{\rm arcsec}$ with a
mean of $\sim1.0\;{\rm arcsec}$ in FWHM, as seen in Figure
\ref{figure:seeing}.

\subsection{Fiber Allocation Procedures}\label{subsection:fiber_allocation}

Fibers were allocated to the selected targets using the software
Spine-to-Object (S2O, version 20101007). S2O takes ``.fld'' files
prepared by users with the coordinate list of science targets as input
and produces ``.s2o'' files as outputs containing the allocation
results.  In addition to science targets, several types of stars need
to be included in the .fld files: guide stars, coordinate calibration
stars (CCS), and flux calibration stars (FCS).  We used the Two-Micron
All-Sky Survey (2MASS) point sources catalog (\cite{Cutri}) with
available NIR magnitudes for these stars. To reduce the systematic
uncertainties about the coordinates given in 2MASS and CFTHLS, we
examined the cross-matched objects between the two, and found
systematic offsets (typically $\sim 0.07$ arcsec, depending on the
four fields of W1--4) between the coordinates reported by these two
surveys. These offsets reduce the number of cross-matched stars, and
hence this offset was corrected for each of the four fields, and the
cross-match was taken again.  The random errors for individual objects
between the two systems are typically $\sim0.2$ arcsec.  We used these
cross-matched objects as the star sample for all FastSound
observations, using the CFHTLS coordinates for them to be consistent
with target galaxies.

FMOS requires at least three guide stars near the edge of the FoV, and
they should be bright in optical bands.  We selected guide stars which
satisfy the following conditions: (i) $11.5<z'<15.4$, (ii)
$11.5<R<15.5$, and (iii) located at $0.22$--$0.30\;{\rm deg}$ from the
center of a FoV. Typically 3--6 guide stars were selected in the final
s2o output. Guide stars are observed by dedicated guide fibers of
FMOS, and they do not affect the number of scientific targets.
Coordinate calibration stars (CCS) were used to correct the rotational
offset of FoV.  We selected $\sim5$ CCSs close to the center of FoV
and $\sim10$ distant from the center.  CCSs in the outer regions
provide an efficient rotational calibration, while CCSs near the FoV
center are used for focusing.  The sky camera takes the image of CCSs
and hence no fibers need to be reserved for these stars.  The
selection criteria for the CCSs were (i) $11.5<z'<15.4$ and (ii)
$11.5<R<15.5$.  Faint stars ($15$--$18$ mag in $JH$ bands) are needed
as FCSs for spectral calibration of science targets. We allocated
about eight fibers to FCSs (i.e., four stars for each of IRS1 and
IRS2), which are the same fibers as those for scientific targets.  The
selection criteria were (i) $0.3 < g'-r' < 0.5$, (ii) $16.5 < r' <
18.0$, and (iii) $H>16.25$.  These conditions were driven by the
desire to select G type stars for calibration as they have flat
spectra and do not have strong absorption lines in $H$ band.

The parameters adopted in the S2O software are as follows.  The
beam-switch offset was set to be $(10,0)\;{\rm arcsec}$ in the Echidna
system coordinates.  The target priority (``1''=highest and
``9''=lowest) was set to ``1'' for guide stars and ``3'' for other
objects.  Another important parameter for fiber allocation is position
angle (PA) of FoV relative to the celestial sphere.  For each FoV, we
tried various PAs to find the best one, for stable guiding (i.e., a
sufficient number of guide stars) and to maximize the number of
science target galaxies.  For each FoV we checked that roughly equal
numbers of FCSs were included for IRS1 and IRS2 fibers, and that the
FCS fibers were not at edge of the detectors, to avoid a decrease of
flux calibration accuracy.  We accepted redundant observations and we
did not omit galaxies at the edge of FoV from our target list, even if
they were already observed by an adjacent FoV.

\begin{figure}
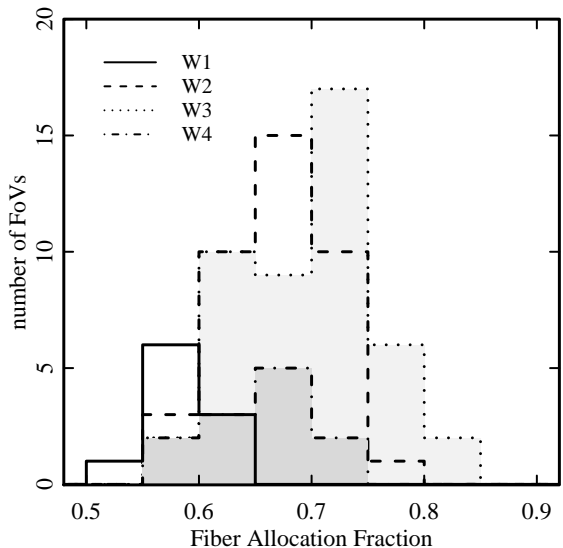

 \begin{center}
 \FigureFile(80mm,80mm){figure5.eps}
 \end{center}
 \caption{The histogram of the fraction of fiber-allocated
galaxies in those selected as the FastSound target galaxies
in \S\ref{subsection:selection}.}
\label{figure:frac_allocated}
\end{figure}

Typically we allocated $\sim360$ fibers to objects, although there are
$\sim400$ fibers and $\sim500$ targets within a FoV, which is mainly
due to the fiber collision and non-uniform distribution of galaxies on
the sky.  This is illustrated in Figure \ref{figure:frac_allocated} as
a histogram of the fiber allocation percentage relative to the number
of target galaxies selected by the FastSound selection condition
(\S\ref{subsection:selection}).  The home positions of FMOS fibers are
in a regular triangular lattice pattern with the fiber spacing of
$84''$, and there is no inaccessible area in the FoV because the
patrol area of the fibers is $87''$ arcsec in radius.  However, the
minimum allowed separation between neighboring fibers is $12''$
arcsec. Therefore spines cannot be allocated to a pair of galaxies if
they are closer than $\sim12''$ arcsec.  This effect will be taken
into account in the clustering analysis reported in the forthcoming
FastSound papers.

\begin{figure}
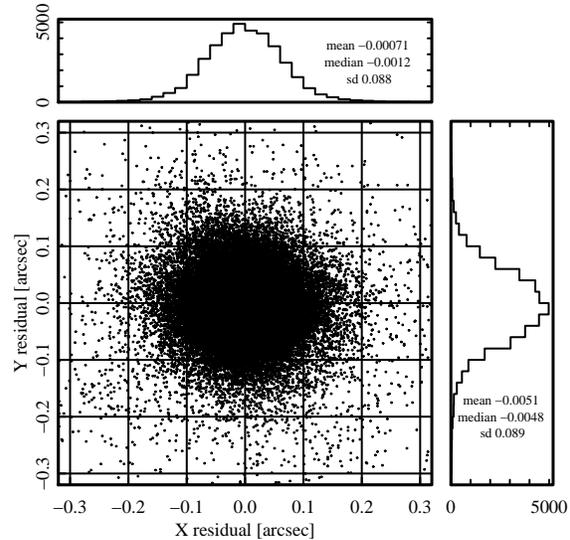

 \begin{center}
 \FigureFile(80mm,80mm){figure6.eps}
 \end{center}
 \caption{The fiber pointing accuracy for the FastSound scientific target
   galaxies, showing the residual between the actual fiber positions
   and input coordinates of targets. 
}\label{figure:residual}
\end{figure}

Figure \ref{figure:residual} shows the accuracy of fiber allocation to
targets, derived as the difference between input fiber positions in
the ``.s2o'' files and those recorded during observations.  The offset
is within $0.12\;{\rm arcsec}$ for almost all targets, which is
consistent with the designed performance of the Echidna system
(\cite{Kimura}).

\subsection{A Note on OH Masks}

The OH mask mirrors are an important feature of FMOS and are used to
suppress the strong OH airglow in the near-infrared bands.  When the
FastSound observations started (Mar. 2012), two different mask mirrors
were used for IRS1 and IRS2, made by different techniques: thin
stainless steel wires were placed at the positions of OH airglow lines
for IRS1, while the mirror of IRS2 was directly coated with reflective
gold. There was also a difference of the masked wavelength regions
between the two; the mask mirror of IRS2 masked fewer OH lines than
that of IRS1, because the former did not cover some faint OH lines.
It was found that the IRS2 mask worked better and was more stable than
IRS1, because the steel metal was sensitive to minute-scale changes of
temperature in the refrigerator.  Therefore the IRS1 mask mirror was
replaced with that masking the same wavelength regions and made by the
same technique as IRS2 in an engineering run conducted in Sep 7--9
2012, i.e., during the 2-year FastSound observation campaign.  This
change will be appropriately taken into account in the galaxy
clustering analysis. The proportion of the masked wavelength region
for IRS1 decreased from $22\%$ to $10\%$ in the whole FMOS
wavelength range in LR, and from $27\%$ to $16\%$ in the HR wavelength
range (H short+) used for FastSound, by this upgrade.

\section{Data Reduction}\label{section:reduction}

\begin{figure}
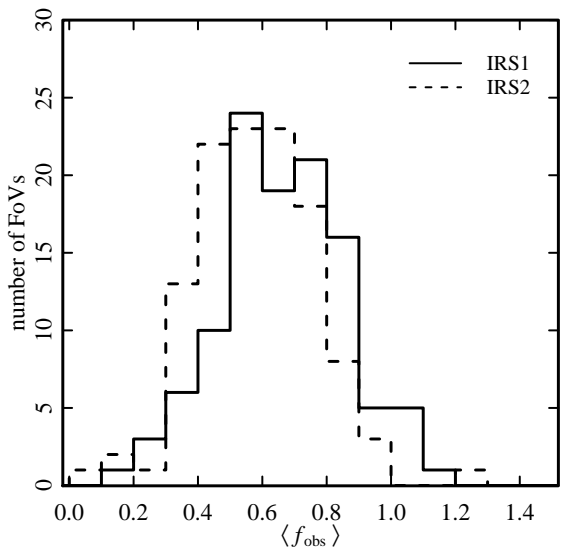

 \begin{center}
 \FigureFile(80mm,80mm){figure7.eps}
 \end{center}
 \caption{The distribution of $\langle f_{\rm obs} \rangle$ within each FoV.}
\label{figure:fobs}
\end{figure}

The FastSound data were reduced with the FMOS data reduction pipeline
(FIBRE-Pac: \cite{Iwamuro}). The software performs
relative flux (i.e., spectral shape) calibration by choosing one of
the flux calibration stars clearly seen on the 2D image in a FoV.  The
line flux reported in the FastSound catalog is calculated from the
observed counts of the final spectral data produced by FIBRE-Pac,
where the counts are normalized so that $1$ count per pixel
corresponds to $1 \mu$Jy.  It should be noted that this absolute flux
normalization is not calibrated by FCSs, but by assuming a fixed total
throughput of $5\%$ (in the HR mode) obtained in good observing
conditions (\cite{Iwamuro}).  Varying observational conditions and the
fiber aperture effect against point sources can be corrected by the
quantity $f_{\rm obs}$, which is defined as the ratio of the flux
calculated by FIBRE-Pac for a FCS to that reported in the 2MASS
catalog.  There is typically $\sim20\%$ scatter in $f_{\rm obs}$ for
individual FCSs within a FoV, caused mainly by fiber positioning
errors and chromatic and/or instrumental aberrations.  The average of
$f_{\rm obs}$ within each FMOS FoV ($\langle f_{\rm obs} \rangle$) was
calculated and included in the FastSound catalog.  In Figure
\ref{figure:fobs}, we show the distribution of $\langle f_{\rm obs}
\rangle$ of all the FastSound FoVs, whose average is $\sim0.6$.  The
total flux from a point source can be estimated by dividing the line
flux in the FastSound catalog by $\langle f_{\rm obs} \rangle$, but
for extended sources, an additional correction for the fiber aperture
effect would be required to estimate the total flux.  The detailed
procedure to convert the uncalibrated flux in the catalog to the
calibrated total flux from emission line galaxies will be described in
Paper II (\cite{Okada}).

Emission lines were searched for automatically using the software,
FMOS Image-based Emission Line Detection (FIELD), developed for the
FastSound project (\cite{Tonegawa2}).  We briefly summarize the
algorithm here.  The continuum component is first subtracted from the
2D spectral image. Then the 2D image is convolved with a 2D Gaussian
kernel that has a similar wavelength and spatial dispersion to the
typical line profile of star forming galaxies. The line
signal-to-noise is calculated by this convolved flux and its error,
and the peaks are searched along the wavelength direction. Special
care needs to be taken to reduce the number of false detections in the
neighborhood of the OH mask regions, which is especially important
for a line search in the FMOS data.  Peaks with $S/N$ larger than a
given threshold are selected as the line candidates. 

In order to avoid multiple detections from one emission line, we set
the minimum separation of candidates to be $20$ pix along the
wavelength direction. If there were more than two $S/N$ peaks within
this separation, we selected the one with the largest $S/N$ as a line
candidate.  The $20$ pix scale corresponds to $\sim 440 {\rm km/s}$,
and therefore emission lines of normal galaxies with a velocity
dispersion of $\lesssim$ 200 km/s should not be blended with
neighboring line candidates.  Finally, we removed a small fraction of
objects from the candidate sample, by using some 2D shape parameters
indicating that they are spurious objects.

\begin{figure}
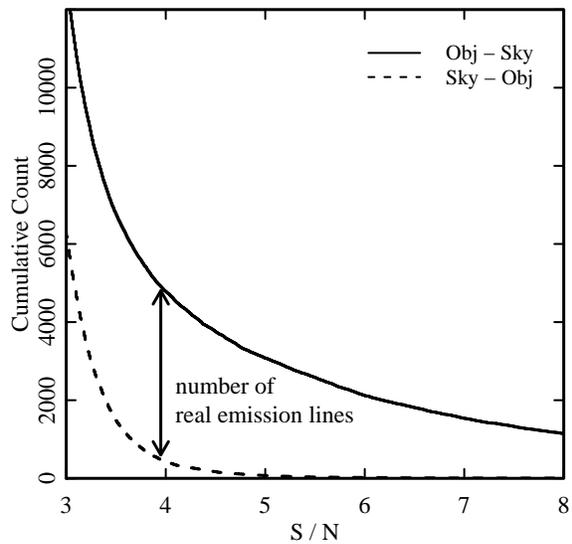

 \begin{center}
 \FigureFile(80mm,80mm){figure8.eps}
 \end{center}
 \caption{The cumulative
   $S/N$ distribution of all the emission line candidates in the
   FastSound survey.  The solid and dashed lines show the number of
   candidates detected in the normal and inverted frames,
   respectively.  The number of real emission lines can be estimated
   by the difference between the two.}
\label{figure:sn}
\end{figure}

In this way we detected $4,797$ emission line candidates at $S/N \geq
4.0$ in all the FastSound data set (see Table \ref{table:linestat} and
Figure \ref{figure:sn} for the dependence on $S/N$).  It should be
noted that the number of galaxies hosting these candidates is $4,119$,
because $\sim200$ galaxies are detected with more than two different
lines (see Paper II, \cite{Okada}), and $\sim400$ galaxies are detected in two
different FastSound FoVs because they were observed twice in the
regions of two overlapping FMOS FoVs. Typically, the number of
detected emission line galaxies in one FMOS FoV is about $10\%$ of the
$400$ fibers.  This relatively low success rate is consistent with
that obtained in our pilot observations reported in a previous paper
(\cite{Tonegawa}). In that paper, we discussed the factors determining
the success rate, by checking how H$\alpha$ emitters found in a
narrow-band survey were omitted from the FMOS targets.  The main
reasons were the uncertainty in the photometric redshift estimates,
which is large relative to the FastSound redshift coverage, and the
uncertainty in the H$\alpha$ estimates.  There are about $380$
H$\alpha$ emitters in one FMOS FoV whose H$\alpha$ fluxes are
detectable by FMOS, and hence a detection efficiency close to $100\%$
should be possible, if the target selection is perfect. However, only
about $1/3$ of them remain after the photo-$z$ cut, and another
$\sim1/3$ remain after the H$\alpha$ flux and color cuts, resulting in
$\sim10\%$ detection efficiency.

All the $4,797$ line candidates cannot be guaranteed to be real
emission lines, as the false line detection rate cannot be reduced to
zero. The false detection rate increases with decreasing threshold of
the line $S/N$.  We estimated the number of false detections by
applying FIELD to inverted images, which are obtained by exchanging
object and sky frames in the reduction process. It is expected that
the rate of false detection of spurious objects should be the same for
the normal (Obj.$-$Sky) and inverted (Sky$-$Obj.)  frames, because the
analysis procedures are exactly the same except for swapping the
object/sky frames.  (Absorption lines may appear as emission lines in
the inverted frames, but the continuum emission is hardly detected for
most of FastSound galaxies and absorption lines are under the
detection limit.)

The numbers of detected line candidates in all the FastSound data in
normal and inverted frames are summarized in Table
\ref{table:linestat} and shown as a cumulative histogram of $S/N$ in
Figure \ref{figure:sn}.  The contamination of spurious objects is
about 10\% above $S/N = 4$, and sharply increases for a lower $S/N$
threshold.  Therefore the number of real emission lines detected by
FastSound can be estimated to be about $4,300$ at $S/N \geq 4$. The $441$
lines at $S/N \geq 4$ in the inverted frames are hosted by $398$
galaxies, and hence the number of galaxies with real emission lines
above this $S/N$ should be $4,119 - 398 \sim 3,700$.  More than 90\%
of these lines are expected to be H$\alpha$ (\cite{Tonegawa}; Paper II).  

\begin{figure*}
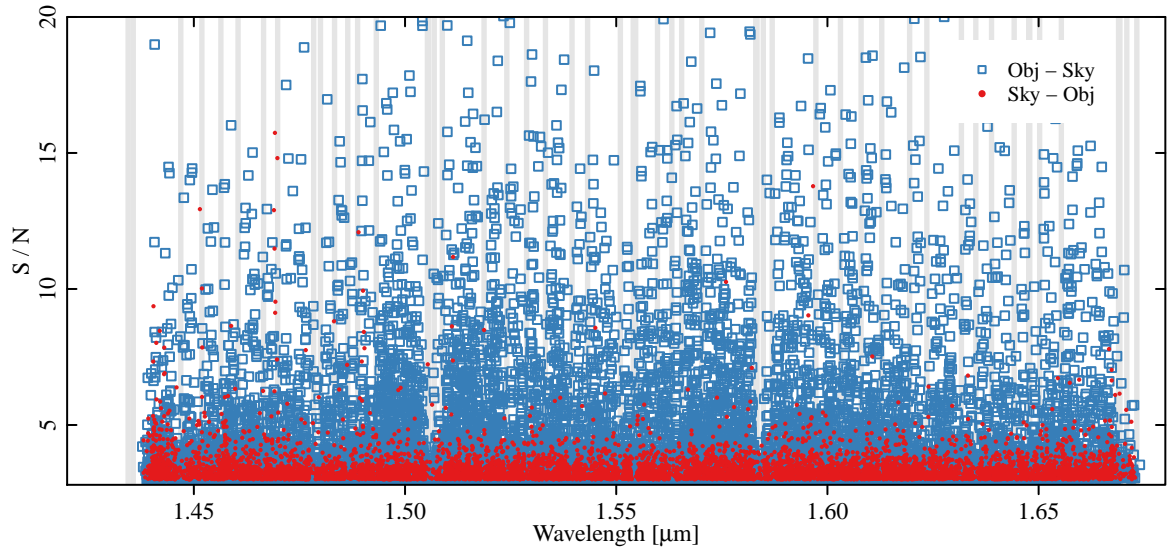

 \begin{center}
 \FigureFile(160mm,80mm){figure9.eps}
 \end{center}
 \caption{ The $S/N$ of emission line plotted against the observed
   wavelength.  Blue and red symbols represent the detected objects in
   normal and inverted frames, respectively.  Grey vertical stripes
   show the positions of OH masks of IRS2. }
\label{figure:wlsn}
\end{figure*}

The wavelength of the detected objects in normal and inverted frames
are presented against $S/N$ in Figure \ref{figure:wlsn}. Here, the
wavelengths were derived by 1D Gaussian fits with a velocity
dispersion as a free parameter for detected lines. When an OH mask
region is overlapping with a line, the line center may be on the mask
region.  The number of fake lines on the inverted frames rapidly
increases below $S/N=5.0$, indicating that one should adopt some $S/N$
threshold not to include spurious objects into analysis.

\begin{table}
\footnotesize
\begin{center}
  \caption{The statistics of the detected line candidates in all the
    FastSound data set.\footnotemark[*] }\label{table:linestat}
\begin{tabular}{cccc}
\hline
\hline
  & Obj.$-$Sky (1) & Sky$-$Obj. (2) & Contamination (2)$/$(1)\\
\hline
$S/N\geq5.0$ & $3,080$ & $72$ & $2.3\%$ \\
$S/N\geq4.5$ & $3,769$ & $170$ & $4.5\%$ \\
$S/N\geq4.0$ & $4,797$ & $441$ & $9.2\%$ \\
$S/N\geq3.5$ & $6,805$ & $1,510$ & $22.2\%$ \\
$S/N\geq3.0$ & $12,795$ & $6,279$ & $49.1\%$ \\
\hline
\hline
\end{tabular}
\end{center}
\footnotemark[*]{Statistics for both the normal (Obj.$-$Sky) and
  inverted (Sky$-$Obj.) images are presented. The inverted frames
  should include only spurious objects. }
\end{table}

\section{Conclusion}\label{section:conclusion}

This is the first in the series of papers based on the FastSound project, a
cosmological galaxy redshift survey designed to detect redshift space
distortions (RSD) in the clustering of galaxies at $z\sim1.4$.  The survey
targets H$\alpha$ emission line galaxies using the near-infrared fiber
multi-object spectrograph (FMOS) of the Subaru Telescope, which has a
circular FoV with $30$ arcmin diameter including $400$ fibers.  The main
scientific goal of the project is to investigate the origin of the accelerated
expansion of the Universe and to test General Relativity using
a measurement of the growth rate of large scale structure.

In this paper, we presented the survey design, the observation
strategy and the basic data reduction adopted by FastSound.  We used
the CFHTLS Wide catalog in the four fields (W1--W4) to select
potential target galaxies, and selected galaxies expected to have high
H$\alpha$ flux based on photometric estimates of redshifts and
star formation rates using the five CFHTLS optical bands.

The observations were carried out over $35$ nights from April 2012 to
July 2014.  We observed $121$ FMOS FoVs in total ($10$, $39$, $54$,
and $18$ for W1--4, respectively), corresponding to a total area of
$20.6\;{\rm deg^2}$ tiled with a continuous hexagonal pattern.  The
data processing was done using the standard FMOS reduction pipeline
(FIBRE-Pac) and the automatic line detection software for FMOS
(FIELD).  We detected $\sim 4,700$ emission line candidates at
$S/N\geq4.0$ corresponding to the line flux sensitivity limit of $\sim
2 \times 10^{-16} \ \rm erg \ cm^{-2} s^{-1}$.  About 10\% of these
should be spurious detections, as judged from the statistics using the
inverted frames.  Removing duplications by multiple lines in a galaxy
and by galaxies observed twice, the number of real emission line
galaxies is estimated to be $\sim 3,700$ at $S/N \geq4$.  More than
$90\%$ of these are considered to be H$\alpha$ emitters at $z \sim
1.2$--$1.5$, and most of the non-H$\alpha$ contaminants are [OIII]
emitters at $z \sim2$ (see Paper II, \cite{Okada}).

The forthcoming papers will report on the properties of detected
emission line galaxies and the FastSound catalog (Paper II, \cite{Okada}),
on metallicity evolution (Paper III, \cite{Yabe}), on
galaxy clustering analyses to derive cosmological constraints (Paper
IV, Okumura et al., in preparation), and other topics such as the
analysis of halo occupation distribution to determine the relationship
between the emission line galaxies and their host halos (Hikage et
al., in preparation).

\bigskip

We are grateful to C. Blake for useful discussions.  The FastSound
project was supported in part by MEXT/JSPS KAKENHI Grant Numbers
19740099, 19035005, 20040005, 22012005, and 23684007.  
KG acknowledges support from Australian Research Council Linkage
International Fellowship grant LX0989763. AB gratefully
acknowledges the hospitality of the Research School of Astronomy \&
Astrophysics at the Australian National University, Mount Stromlo,
Canberra where some of this work was done under the Distinguished
Visitor scheme. SM and TO were supported by World Premier
International Research Center Initiative (WPI Initiative), MEXT, by
the FIRST program ``Subaru Measurements of Images and Redshifts
(SuMIRe)'', CSTP.

\end{document}